\documentclass[prd,superscriptaddress,onecolumn,showpacs,nofootinbib]{revtex4}

\usepackage{graphicx}
\usepackage{dcolumn}
\usepackage{bm}
\def\bea{\begin{eqnarray}}
\def\eea{\end{eqnarray}}

\newcommand{\nn}{\nonumber}
\newcommand{\na}{\nabla}
\def\beq{\begin{equation}}
\def\eeq{\end{equation}}

\def\pa{\partial}

\def\na{\nabla}

\def\M{\mu}
\def\N{\nu}

\begin{document}

\preprint{APS/123-QED}

\title{Dark spinor model with torsion and cosmology}
\author{Seyen Kouwn}\email{seyen@skku.edu}
\affiliation{Department of Physics and Institute of Basic Science,
 Sungkyunkwan University, Suwon 440-746 Korea}

\author{Joohan Lee}
\email{joohan@kerr.uos.ac.kr} \affiliation{Department of Physics,
University of Seoul, Seoul 130-743 Korea}
\author{Tae Hoon Lee}\email{thlee@ssu.ac.kr}
\affiliation{Department of Physics and Institute of Natural
Sciences,\\ Soongsil University, Seoul 156-743 Korea}
\author{Phillial Oh}
\email{ploh@newton.skku.ac.kr} \affiliation{Department of Physics
and Institute of Basic Science, Sungkyunkwan University, Suwon
440-746 Korea}

\date{\today}

\begin{abstract}
We study cosmological consequences of the dark spinor model when  torsion is included. Only some components of the torsion are allowed to be non-vanishing in homogeneous and isotropic cosmology, but there exist freedoms in the choice of these components which is consistent with the evolution equations. We exploit this and discuss several cases which can result in interesting cosmological consequences.
Especially, we show that there exist  exact cosmological solutions in which the Universe began its acceleration only recently   and  this solution is an attractor. This corresponds to a specific form of the torsion with a mild  fine-tuning which can address the coincidence problem.
\end{abstract}


\maketitle

\section{Introduction}



To search for the theoretical foundation of the current accelerating Universe
\cite{ob} has become one of the most fundamental problems in modern cosmology. An unknown source of energy which is repulsive in nature referred to as the dark energy is one of the most promising candidates and various proposals for its origin have been put forward  \cite{amen}.  But
so far, relatively little attention has been  paid to the possibility
of fermionic sources in cosmology even though it has been known for some time that  spinor fields could play some  important roles in the evolution of the Universe \cite{ferm,picon,riba,saha}.
  For example, in Ref. \cite{picon}  it is shown that a spinor field can accommodate any desired behavior of its energy density if
an appropriate  self-interaction of the spinor field is introduced and it is suggested that it can account for the dark
energy and allows for realistic cyclic non-singular solutions. In Ref. \cite{riba} it is investigated whether fermionic sources could be responsible for the accelerating periods during
the evolution of a universe after a matter field would guide for the decelerating period. In Ref. \cite{saha} the authors have shown
that it is possible to simulate perfect fluid and dark energy by means of a nonlinear spinor field.

Recently, the so-called ELKO spinor,
which is a new spin-half field  defined as the eigenspinor of the charge conjugation operator with mass dimension one, was introduced in Ref.  \cite{grin}.
They have many special properties, one being that  their
dominant coupling is via the gravitational field and therefore
are naturally dark.
It  has been proposed to be a candidate of dark matter \cite{grin} and dark energy \cite{burnett} and many
cosmological aspects of the ELKO
spinor model have been investigated \cite{many}. In addition,
 dark spinor was
shown to be a potential candidate to probe non-trivial topologies in
spacetime \cite{1} and to be incorporated in a natural extension of the
Standard Model by representing a mass dimension-transmuting operator
between Dirac and ELKO spinor \cite{2}.

In Ref. \cite{bohmer}, however, it was pointed out that the construction of ELKO spinors
itself implicitly violates Lorentz invariance. The authors have proposed a non-local but
Lorentz invariant version of the dark spinors. Also it is pointed out that
some crucial errors of neglecting the spin connection part of the fermionic sector in the computations of stress energy tensors  were made in the previous literatures on ELKO models. The correct expression for the ELKO spinor field was applied to dark energy and the result showed the existence of the de Sitter acceleration of the Universe.
Another interesting consequences of dark spinor in cosmological aspects is that the equation of the state $\omega$ for the
dark spinor 
can be negative  and can cross the phantom-divide even without any  phantom-like matter carrying negative kinetic energy
  \cite{cald}. It  converges to a parameter value within the bound produced by using WMAP data \cite{hins} as was discussed in Ref. \cite{burnett}.

In spite of these nice features, further dynamical analysis of the ELKO model
following this work using the correct form of the stress energy momentum
tensor revealed that scaling attractor solutions do not exist in this
model \cite{wei}, and it has the difficulty of addressing the coincidence problem. This aspect of the dark spinor model draws our attention and motivates to extend to the case in which  torsion \cite{hamm} is included in the theory. First, recall that when the cosmological principle is extended to  the general relativity theory with torsion,  only the time component of the trace and totally anti-symmetric components of the torsion tensor can be nonzero \cite{tsam}. For ordinary Dirac spinor, it is well-known that  it interacts only with the totally anti-symmetric components of the torsion tensor and
the spin connection part does not contribute to the total energy density \cite{hehl1}. On the other hand, the dark spinor interacts with all components of the torsion and we investigate the cosmological consequence of the torsion in the dark spinor model. In particular, we find the time component of the non-vanishing trace vector could provide a novel feature of explaining the coincidence problem. That is,  with a mild fine-tuning of the parameters, we show that there exist a cosmological solution in which the Universe began its acceleration only recently   and  that the solution is an attractor.

The paper is organized as follows: In Sec. \textup{II}, we give a brief summary of the spinor field in curved background with torsion in order to set the notation. In Sec.  \textup{III}, the action of the dark spinor is given and the equations of motions are calculated. In Sec.  \textup{IV}, we discuss various cosmological solutions focusing on the solution which can address the coincidence problem and stability analysis of this solution is performed. Sec. \textup{VI} contains conclusion and discussions.



\section{spinor in curved background}
In this section, we give a brief summary of  the spinor field in curved background which is essential for our analysis.
More detailed descriptions can be found in Ref. \cite{spinref}.
The gamma matrices are given by
\bea
\gamma^0= \left(
           \begin{array}{cc}
             0 & 1 \\
             1 & 0 \\
           \end{array}
         \right),~~\gamma^i=\left(
                              \begin{array}{cc}
                                0 & \sigma_i \\
                                \ -\sigma_i & 0 \\
                              \end{array}
                              \right)
                           \eea
 with
         $\{\gamma^a, \gamma^b\}=-2\eta^{a b},~~ \eta^{a b}= diag.(-1,1,1,1)$

We introduce the metric and tetrad via
\bea
g_{\mu\nu}=e^a_\mu e^b_\nu \eta_{ab}.
\eea
Imposing the metricity $\nabla_\mu g_{\alpha\beta}=0,$ the connection is given by
\bea
\Gamma^\rho_{\mu\nu}=
\{\begin{array}{c}
  \rho \\
    \mu\nu
\end{array}\}
 -K^{~~~\rho}_{\mu\nu},
\eea
where $\{ \}$ is the Christoffel symbol and the contortion $K^{~~~\rho}_{\mu\nu}$ is given by the torsion tensor
$S^{~~~\rho}_{\mu\nu}=\Gamma^\rho_{[\mu\nu]}$ via \cite{hamm}
\bea
K^{~~~\rho}_{\mu\nu}= -\left(S^{~~~\rho}_{\mu\nu}+S^{\rho}_{~\mu\nu}+
S^{\rho}_{~\nu\mu}\right).\label{contorsin}
\eea

The covariant derivative of the spinor and
its dual are given by
\beq \na_{\M}\psi \equiv \pa_{\M}\psi-\Gamma_{\M}{\psi} \eeq and
\beq \na_{\M}{\bar\psi}
\equiv\pa_{\M}\bar{\psi}+\bar{\psi}\Gamma_{\M}, \eeq
where $\Gamma_{\M}$ is the connection on the spinor.

We introduce the spin connection by imposing the metricity for the tetrad
\bea
0=\nabla_\mu e^a_\nu = \partial_\mu e^a_\nu -\Gamma^\rho_{\mu\nu} e^a_\rho + \omega_{\mu~b}^{~a} e^b_\nu,
\label{spinconnect}
\eea
where the spin connection satisfy $\omega_{\mu~b}^{~a}=-\omega_{\mu b}^{~~~a}$ and
\bea
\omega_{\mu}^{~ab} = e^a_\nu\left( \partial_\mu e^{b\nu} + \Gamma^\nu_{\mu\rho}  e^{b\rho}\right).\label{spinconnection}
\eea
We also impose the covariant constancy of the gamma matrices;
\bea
0=\nabla_\mu \gamma^\rho = \partial_\mu \gamma^\rho +\Gamma^\rho_{\mu\nu} \gamma^\nu - [\Gamma_\mu, \gamma^\rho].\label{gammamtx}
\eea
From the eqs (\ref{spinconnect}) and (\ref{gammamtx}), we obtain
\bea
\Gamma_\mu = \frac{1}{4} \omega_{\mu}^{~ab} \gamma_a \gamma_b + c I.\label{spinorconnection}
\eea
We consider only the case when the constant $c$ is equal to zero.

We can decompose the contortion tensor (\ref{contorsin}) into
 a traceless part and trace \cite{sa}:
\begin{equation}
{K_{\mu\nu}}^{\rho} = {\tilde{K}_{\mu\nu}}^{~~~\rho} -
\frac{2}{3}\left( \delta_{\mu}^{\rho} S_\nu  -
g_{\mu\nu} S^\rho \right), \label{decompos}
\end{equation}
where ${\tilde{K}_{\mu\nu}}^{~~~\rho}$ is the traceless part,
${\tilde{K}_{\mu\nu}}^{~~~\mu}=0$
 and $S_\nu$ is
the trace of the torsion tensor, $S_\nu = S^{~\ \
\mu}_{\mu\nu}$.
Making use of  (\ref{decompos}),
  we can write curvature scalar as follows:
\begin{eqnarray}
\bar{R} = R - 4{\nabla}_\mu S^\mu -\frac{8}{3}S_\mu S^\mu -
\tilde{K}_{\nu\rho\alpha} \tilde{K}^{\alpha\nu\rho}.\label{r2}
\end{eqnarray}
In the right hand side of the above equation, scalar curvature and the covariant derivative are calculated with respect to the Christoffel symbol.

\section{Torsion dark spinor model}

Let us
consider the dark spinor action of the form
\bea S=\int d^{4}x \sqrt{-g} \left[
 \frac{\bar R}{2\kappa}
 -\frac{1}{2}g^{\M\N}\na_{\M}\bar{\psi}\na_{\N}\psi-V(\bar{\psi}\psi) \right]
+S_{m}, \nn  \eea
where  $\bar R$ is the scalar curvature including the torsion piece and $V$ is the potential for the dark spinor (ELKO) field $\psi$ and its dual $\bar{\psi}$. $S_m$ is the matter field action.

To discuss the Friedman cosmology in the flat Robertson-Walker space-time,
consider a metric of the form
\beq ds^2=-N^2 dt^2+ a^2(t) dx^i dx^i,\label{RW}\eeq
where $a(t)$ is the scale factor
of our three dimensional universe. We assume
\beq
\psi(x^\M)=\phi(t)~ \xi \label{elko} \eeq with a homogeneous real
scalar field $\phi(t)$ and a constant spinor $\xi$ such that $\bar\xi
\xi=1$ but $\na_\M \xi \ne 0$.
It is to be pointed out that that the spinor nature of the dark spinor is remnantal  in the
real scalar field $\phi(t) $ defined in (\ref{elko})  and the ensuing evolution equations
 contain the contributions from the spin connection which are absent in the case of
a genuine scalar field.
 It was shown  that these new contributions supply
fundamentally different aspects in the dark spinor cosmology \cite{bohmer}.



For isotropy and homogeneity of Universe, one assumes the following non-vanishing components
of the torsion \cite{tsam}
\bea
S^{~~~1}_{10}=S^{~~~2}_{20}=S^{~~~3}_{30}=h(t)/2,~ S^{~~i}_{jk}=S^{~~~i]}_{[jk}=f(t)/3\epsilon_{jk}^{~~i}.\label{torsion}
\eea
Eqs. (\ref{RW}) and (\ref{torsion}) give the following connection components:
\bea
&&\Gamma^0_{00}= \frac{\dot N}{N},~ \Gamma^0_{ij}=\frac{ab}{N^2}\delta_{ij},~ \Gamma^i_{j0}=\frac{b}{a}\delta_{ij},\nonumber\\
&&\Gamma^i_{0j}=\frac{\dot a}{a}\delta_{ij},~ \Gamma^i_{jk}=f\epsilon_{ijk},\label{connection}
\eea
with $b = \dot a + ah.$
Using the above components, we obtain the scalar curvature
\bea
\bar R= 6\big[ \frac{1}{aN} \frac{d}{dt} \left( \frac{b}{N}\right) + \left(\frac{b}{a N}\right)^2-\left( \frac{f}{a}\right)^2\big].\label{cub}
\eea
Note that the above result (\ref{cub}) agrees with Eq. (\ref{r2}) up to a total derivative term. From the metric (\ref{RW}), the tetrad is given by
\bea
e^a_\mu=(N, a, a, a),~~ e_a^\mu = (\frac{1}{N}, \frac{1}{a}, \frac{1}{a}, \frac{1}{a}).
\eea
Using (\ref{connection}), the only non-vanishing components of the spin connection (\ref{spinconnection})
are given by
\bea
\omega_{x^i~0}^{~~j}=\omega_{x^i~j}^{~~0}=\frac{b}{N}\delta_{ij},~\omega_{x^i~k}^{~~j}=f\epsilon_{ijk}.
\eea
Then, we have from eq.(\ref{spinorconnection})
\bea
\Gamma_0=0,~~\Gamma_{x^i}= \frac{b}{2N}\gamma_0\gamma_i + \frac{f}{4}\epsilon_{ikj}\gamma_j\gamma_k.
\eea

Putting everything together, we have ($\kappa=1)$
\bea
 S=\int dt  \Big[ \frac{1}{N} \left( -3a {\dot a}^2 + 3
 a^3h^2+\frac{1}{2} a^3\dot\phi^2+\frac{3}{8}a b^2\phi^2\right)
-N\left( 3 a f^2 +\frac{3}{8}a f^2 \phi^2 + a^3V\right)\Big] +S_m
\eea
The  equation related with the energy density coming from the variation with respect to $N$ is given by
\bea
3H^2 =   \frac{1}{2}  {\dot\phi}^2 + V
+\frac{3}{8} H^2 \phi^2 +\frac{3}{4}  H h ~\phi^2+3\left(1+\frac{1}{8}\phi^2\right) h^2+
3\left(1+\frac{1}{8}\phi^2\right)\frac{f^2}{a^2}+\rho_m,\label{hequation}
\eea
and the one related with the pressure  coming from the variation with respect to $a$ is
\bea
-2\dot{H}-3H^2 =\frac{1}{2}  {\dot\phi}^2-V -\frac{3}{8} H^2 \phi^2-\frac{1}{4}
\frac{d}{dt}\left[\left( H+h \right) \phi^2\right]
+3\left(1+\frac{1}{8}\phi^2\right)h^2 +3\left(1+\frac{1}{8}\phi^2\right)
\frac{f^2}{a^2}+p_m.\label{pequation}
\eea
The scalar field equation is given by
\bea
0=\ddot \phi + 3H\dot\phi + V^\prime (\phi)-\frac{3}{4}
\left( (H+h)^2 -\frac{f^2}{a^2}\right)\phi.\label{scalar}
\eea

Note that the above equations (\ref{hequation}), (\ref{scalar}), (\ref{pequation}) are different from the equations one would obtain
by directly using $\bar G_{\mu\nu}=T^{m}_{\mu\nu}$, where $\bar G_{\mu\nu}$ is the Einstein tensor calculated with the connections including the torsion as in
Eq. (\ref{connection}) \cite{bohm,tilq}. The discrepancy comes from the fact that when torsion is included, the variation $\delta \bar R_{\mu\nu}$ of the Ricci tensor
in calculating the equations of motion from the scalar curvature  is not a total derivative and cannot be neglected as in the case without torsion.
Therefore,  $\bar G_{\mu\nu}=T^{m}_{\mu\nu}$ is not justified in the case when torsion is included.
Differentiation Eq. (\ref{hequation}) with respect to time and using Eqs. (\ref{scalar}) and (\ref{pequation}), we obtain the following
consistency equation:
\bea
0=3h\left[ 2 \dot{h} +6Hh+\frac{1}{2}\left(H+h\right)\phi\dot{\phi}+\frac{1}{4}\left
(\dot{H}+\dot{h}\right)\phi^2+\frac{3}{4}H\left(H+h\right)\phi^2\right]
+6\frac{ f \dot{f}}{a^2}\left(1+ \frac{1}{8}\phi^2\right)
+\dot\rho_m+3H(\rho_m+p_m),\label{consist}
\eea
which serves as a dynamical equations for the torsion.

\section{Cosmological Solutions}
In this section, we discuss diverse cosmological solutions depending on the torsion.
\subsection{Torsion as a dark matter candidate}
We first analyze the pure torsion case where the dark spinor field is absent, $\phi\equiv0$.
Let us also  assume that $\rho_m=p_m=0.$
If we define $\rho_h=3h^2$ and $\rho_f=3 f^2/a^2$, we have from Eqs. (\ref{hequation}) and (\ref{pequation})
\bea
3H^2 =\rho_h+\rho_f, ~~-2\dot{H}-3H^2 =\rho_h -\frac{1}{3}\rho_f.
\eea
The second equation says that $p_h=\rho_h$, whereas
$p_f=-\rho_f/3.$
Eq. (\ref{consist}) gives the continuity equation for the torsion energy density:
\bea
\dot\rho_h + 6H\rho_h+ \dot\rho_f + 2H\rho_f=0.\label{total}
\eea
Let us suppose that $\rho_h$ and $\rho_f$ are separately conserved. Then, $\rho_h\propto 1/a^6$ corresponding to the equation of state $\omega_h=1$
and $f=f_0$ is a constant corresponding to $\omega_f=-1/3$ as expected. However, if we assume $\rho_f=r\rho_h$ with a constant $r>0$ and $\rho_t=\rho_h +\rho_f
=(1+r)\rho_h$,
and the pressure $p_t=\rho_h-\rho_f/3=(1-r/3)\rho_h$,
Eq. (\ref{total}) becomes
\bea
\dot\rho_t + 3AH\rho_t=0,~~ A= \frac{6+2r}{3+3r}.
\eea
Therefore, $\rho_t$ can describe an ideal fluid with the range of the equation of state given by $-1/3\leq\omega_t=\frac{3-r}{3+3r}\leq 1.$
When $r=3$, it describes the cold dark matter. It is to be commented that the relation $\rho_f=r\rho_h$ is very ad hoc, nevertheless nothing
forces  separate conservation as far as continuity equation (\ref{total}) is concerned.

\subsection{Torsion and cosmological constant}
Next, we consider the cases $f=0$.
Let us define a quantity
\bea
X=2h+\frac{1}{4}\left( H+h\right)\phi^2.
\eea
We assume that there is no interaction between the matter and spinor dark energy with torsion, and the matter and the dark spinor are separately conserved. 
\bea
\dot\rho_m+3H(\rho_m+p_m)=0.
\eea
Then from Eq. (\ref{consist}), we have
\bea
3h\left(\dot X + 3HX\right)=0.\label{xxequation}
\eea
We consider $h\neq 0$ and  then, Eq. (\ref{xxequation}) can be immediately integrated to give
\bea
X=\frac{c_1}{a^3}.\nonumber
\eea
Eq. (\ref{pequation}) by using (\ref{hequation}) becomes
\bea
-2 \left(\dot H + \dot h\right) = \dot\phi^2 + 3 (H+h)X+\rho_m+p_m.\label{hprime}
\eea
Also, Eqs. (\ref{hequation}) and (\ref{scalar}) can be rewritten as
\bea
3H^2 =   \frac{1}{2}  {\dot\phi}^2 + V
+\frac{6}{\phi^2} \left(X-2h\right)^2+3 h^2 +\rho_m,\label{prime}
\eea
and
\bea
0=\ddot \phi + 3H\dot\phi + V^\prime (\phi)-12 \frac{\left(X-2h\right)^2}
{\phi^3}, ~~(\prime\equiv \frac{d}{d\phi})\label{hscalar}
\eea
We have four equations (\ref{xxequation}),(\ref{hprime}), (\ref{prime}),
and (\ref{hscalar}). These equations are not independent, and we have three equations to be solved including Eq. (\ref{prime}).

The torsion component $h$ which was introduced for homogeneity and isotropy via (\ref{torsion})
is an external input as long as it satisfies the evolution equations. We exploit this arbitrariness to assume an existence of the functional dependence of $h(t)\equiv h(\phi)$
and  discuss cases where some analytic properties of the above equations can be investigated.

Let us  discuss the case where $\rho_m=p_m=0.$ Under the further simplifying assumption of $c_1=0$, some exact cosmological solutions can be obtained
in a couple of cases which describe the accelerating Universe.
First, $X=0$ and we have
\bea
H=-h\left(1+\frac{8}{\phi^2}\right).\label{hubble}
\eea
We are interested in the case of $h<0$. From Eq. (\ref{hprime}), we have
\bea
\dot\phi = \frac{16}{\phi^3}\left( h_{,\phi}\phi- 2h\right),\label{cosco}
\eea
which yields from Eq. (\ref{prime}) the potential of the form
\bea
V= \frac{24 h^2}{\phi^4}(8+\phi^2)- \frac{128}{\phi^6}\left( h_{,\phi}\phi- 2h\right)^2.\label{pot1}
\eea
One can check that Eq. (\ref{hscalar}) is satisfied with Eqs. (\ref{hubble}), (\ref{cosco}) and (\ref{pot1}).

Let us discuss a couple of cases where exact cosmological solution  can be obtained.
The first case is when the potential is of the form
\bea
V(\phi)=\frac{1}{2}m^2\phi^2+V_0.\label{pott}
\eea
It corresponds to the choice of $h\propto \phi^2$. Comparing with the potential (\ref{pot1}), we
find they coincide if we choose $h=\pm m\phi^2/\sqrt{48}$. Moreover, the cosmological constant
$V_0$ is given by
\bea
V_0= 4 m^2.\label{massy}
\eea
One can check that Eq. (\ref{hscalar}) is also satisfied.
From Eq. (\ref{cosco}), we have $\dot\phi=0$, and $\phi=\phi_0$.
Choosing $h=- m\phi^2/\sqrt{48}$, the Hubble parameter is given by
\bea
H=\frac{m}{\sqrt{48}}\left(8+\phi_0^2\right),
\eea
for the de Sitter accelerating solution.
A couple of comments are in order: First, $\phi$ need not be the minimum of the potential
(\ref{pott}) for its stable equilibrium point and can stay away from zero.
This can be seen dynamically from Eq. (\ref{hscalar}) as follows:
The restoring force coming from $V^\prime$ is exactly canceled by the repulsive
force coming from the torsion if $h\propto \phi^2$. Therefore, the scalar field undergoes
a damped motion without any force acting, and this solution corresponds to the solution in which
if it takes initial value $\phi_0$, it stays there forever.
Also, we note the difference with an ordinary quintessence model with mass term and
cosmological constant. In this case, the $\phi=0$ is dynamically preferred, and the acceleration
can be driven by the cosmological constant. However, the crucial difference is that
there is a priori no relation between the cosmological constant and the mass of the scalar
field, whereas in the dark spinor model with torsion, there is a relation given by
eq. (\ref{massy}). This implies that the smallness of the  cosmological constant is directly
interwoven with the ultra light scalar field.
\subsection{Torsion and coincidence problem}
Another case of interest is when $h$ is given by
\bea
h =-\vert c \vert \exp (-\lambda \phi/2)\phi^2~ (\lambda>0). \label{htr}
\eea In this case,
we have potential of the form
\bea
V=V_*e^{-\lambda \phi}\left[ \phi^2 + c_*\right],~~ V_*= 24c^2,~ c_*=8-\frac{3}{2}\lambda^2.
\label{ppp}
\eea
Eq. (\ref{cosco})  gives
\bea
\dot\phi= \lambda_* e^{-\lambda \phi/2}~~(\lambda_*=8\vert c\vert\lambda),
\eea
which can be integrated to yield
\bea
\phi(t)=\frac{2}{\lambda}\ln t + A.\label{solution}
\eea
We choose the integration constant $A$ of the above equation to be  zero which fixes a relation $4\vert c\vert\lambda^2=1.$
The Hubble parameter $H$ by using Eq. (\ref{hubble}) and its first time derivative are given by
\bea
H=\frac{2}{\lambda^2 t}\left[ 1+\frac{(\ln t)^2}{2\lambda^2}\right],
~~ \dot H=-\frac{2}{\lambda^2 t^2}\left[ 1-\frac{\ln t}{\lambda^2}+\frac{(\ln t)^2}{2\lambda^2}\right].
\eea
To calculate the scale factor, let us define $H_\phi=d \ln a/d \phi$. Then,  Eq. (\ref{hubble}) can be readily integrated to give
\bea
a(\phi)=a_*e^{\frac{1}{\lambda}(\phi + \frac{\phi^3}{24})}
=a_{*}e^{\frac{2}{\lambda^2}((\ln t) + \frac{(\ln t)^3}{6\lambda^2})},
\label{scalee}
\eea

To study the existence of the accelerating solution, we first check
\bea
\frac{\ddot a}{a}=\dot H+H^2=\frac{1}{ t^2}
\left[  -\frac{2}{\lambda^2}\left(1-\frac{2}{\lambda^2}\right)
+2\frac{\ln t}{\lambda^4}-\left(1- \frac{4}{\lambda^2}\right)\frac{(\ln t)^2}
{\lambda^4}+\frac{(\ln t)^4}{\lambda^8}    \right].\label{acta}
\eea
We see that at early times when $\lambda^{3/2}<\ln t<\lambda^2~ (\lambda>1)$,  Eq. (\ref{acta}) describes a decelerating universe. This is the epoch when the first term in $\dot H$ is dominant.
When $\ln t>\lambda^2 $ the $(\ln t)^4$ term begins to become dominant gradually and at  later times , $H^2$ term becomes dominant. The acceleration would begin around $\ln t_{acc}\sim \lambda^2$.
If we adjust the parameter $\lambda$ to be  $\sim 12$ we get the late time acceleration occurring  around $t_{acc} \sim 10^{61} t_{pl }$, which roughly corresponds to the current age of the Universe.
This could explain the coincidence problem.
The dependence of $t_{acc}$ on the numerical value of $\lambda$ is depicted in Fig. 1. Comparing with the $\Lambda-CDM$ which requires a fine-tuning of the order of   $\sim 10^{-120}$ for the cosmological
constant,  it is of
$O(1)$ and poses no fine-tuning issue.
\begin{figure}[ht]
\begin{center}
\scalebox{0.65}[0.65]{ \includegraphics{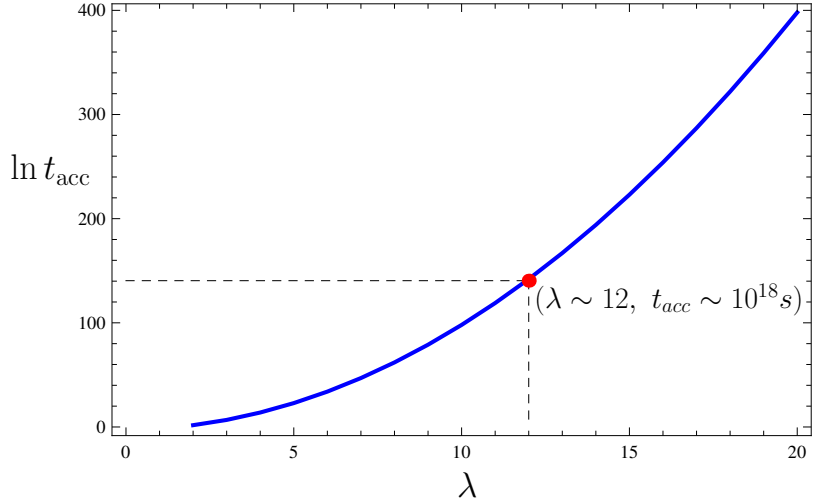} }
\end{center}
\caption{\small
The relation between $t_{acc}$ and the parameter $\lambda$  in the Planck unit. The current acceleration can be achieved with $\lambda \approx 12.$
}
\end{figure}

We perform a stability analysis of the solution (\ref {solution}) and (\ref{scalee}) with $A=0$.
Variation of the Eq. (\ref{hscalar}) with $X=0$ gives
\bea
0=\delta \ddot \phi + 3H\delta \dot\phi+3\dot\phi \delta H + V^{\prime\prime} (\phi)\delta\phi+144 \frac{h^2}{\phi^4}\delta\phi-96\frac{h}{\phi^3}h^\prime\delta \phi. \label{pert}
\eea
$\delta H$ can be eliminated through Eq. (\ref{hubble}) via
\bea
\delta H=\left[-h^\prime(1+\frac{8}{\phi^2})+ 16 \frac{h}{\phi^3}\right]\delta\phi
\label{hubb1}
\eea
Using Eqs. (\ref{htr}), (\ref{ppp}), and (\ref{solution}), we find that Eq. (\ref{pert}) reduces to
\bea
\delta \ddot \phi + \frac{3}{\lambda^4t}\left (2 \lambda^{2}+(\ln t)^2\right)\delta \dot\phi+
\frac{3}{\lambda^4t^2}\left(f(\lambda)+(\ln t)^2\right)\delta\phi=0,\label{pert2}
\eea
with $f(\lambda)$ being
\bea
f(\lambda)= -\frac{3}{4}\lambda^4+2 \lambda^2-\frac{1}{3}.
\eea
A numerical analysis of   Eq. (\ref{pert2}) shows the attractor behavior of the solution, as is shown in Fig. 2.

\begin{figure}[ht]
\begin{center}
\scalebox{0.65}[0.65]{ \includegraphics{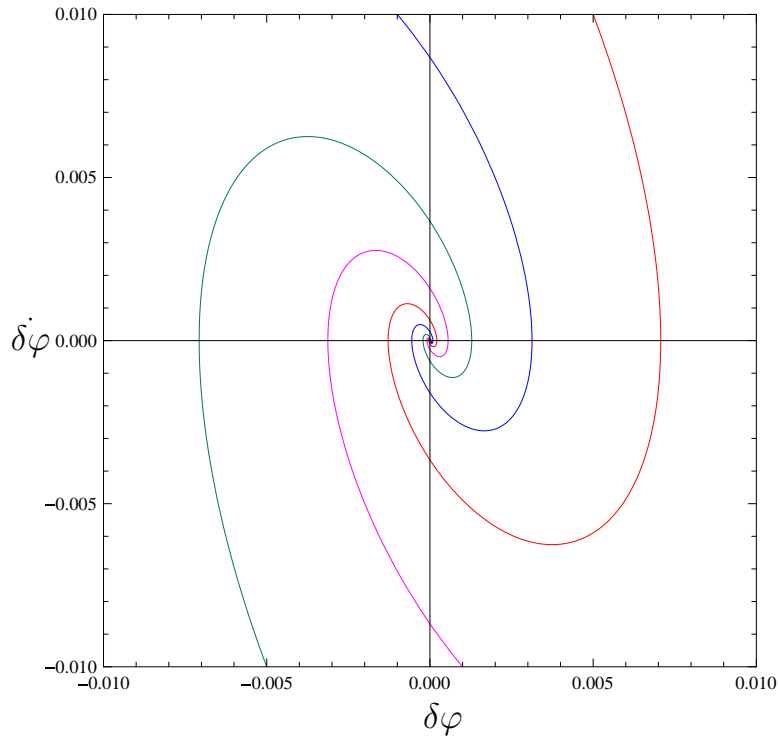} }
\end{center}
\caption{\small
Attractor behavior: The initial condition is given at $N_i \equiv\ln t_i\sim 130$ which corresponds approximately to the decoupling time. The red curve is with $\phi(N_i)=0.005 \,, \dot{\phi}(N_i)=0.01$, blue  with  $\phi(N_i)=-0.001 \,, \dot{\phi}(N_i)=0.01 $, green with $\phi(N_i)=-0.005 \,, \dot{\phi}(N_i)=-0.01 $, and pink : $\phi(N_i)=0.001 \,, \dot{\phi}(N_i)=-0.01 $.
}
\end{figure}

A dynamical estimate how the solution becomes an attractor can be given.
To check the stability in  the asymptotic region where $\ln t>> \lambda^2$, $\phi>> 2\lambda,$ we first note that  from Eq. (\ref{ppp})
\bea
V^{\prime\prime}\delta\phi\sim 24 c^2e^{-\lambda \phi}\phi^2=\frac{6}{\lambda^4t^2}(\ln t)^2\delta\phi, \label{attt}
\eea
and among the last three terms of Eq. (\ref{pert}), $ V^{\prime\prime} (\phi) $ term
dominates. This is a positive quantity and provides a restoring force in the asymptotic region.
Also, from Eq. (\ref{hubb1}), we have $\delta H\sim -h^\prime\delta\phi$, which yields using Eq. (\ref{htr})
\bea
3\delta H\dot\phi\sim -\frac{\lambda}{2} H \delta \phi=-\frac{3}{\lambda^4t^2}(\ln t)^2\delta\phi.
\eea
This term acts as a source of  a repulsive force, but their magnitude is smaller then the attractive force of Eq. (\ref{attt})
Over all, in the asymptotic region, the perturbation equation becomes
\bea
\delta \ddot \phi + \frac{3}{\lambda^4}\frac{(\ln t)^2}{t}\delta \dot\phi+
\frac{3}{\lambda^4}\frac{(\ln t)^2}{t^2}\delta\phi=0.
\eea
It can be easily checked that the perturbation decays as
\bea
\delta \phi\sim \frac{(\ln t)^\alpha}{t},
\eea
and the solution is an attractor in the asymptotic region.

\section{conclusion}

We have examined the possible cosmological role that the torsion could play in the dark spinor model.
Freedom exists in  the choice of  the torsion, which is consistent with the continuity equation. This has been exploited for the choice of diverse forms of the torsion, which resulted in the various cosmological consequences.
 In the pure torsion case without dark spinor, torsion could play the role of the dark matter.
When the torsion is proportional to $\phi^2$, we have an accelerating de Sitter Universe.  In particular, we found an  exact cosmological solution in which the Universe began its acceleration only recently   and  it is an attractor. This corresponds to a specific form of the torsion with a mild  fine-tuning and it can address the coincidence problem.

In this work, the torsion was assumed to be a kind of external matter field whose
dynamics is dictated by the continuity equation.
 Even though this does not confront with any theoretical inconsistency,
their choices were restrictive to certain forms and could be regarded as rather ad hoc.
In the more fundamental approach like the
Einstein-Cartan-Sciama-Kibble theory of gravity \cite{hamm} where the torsion tensor
is regarded as a dynamical
variable from the beginning, the
equations of motion coming from the variation with respect to the torsion generates a spin-spin interaction equation. In the Dirac case, this interaction is significant only at extremely high densities and it was shown that such an interaction averts the unphysical big-bang singularity, replacing it with a cusp-like bounce at a finite minimum scale factor \cite{niko}.
This approach could be extended to the ELKO case \cite{Boehmer:2006qq}. In Ref. \cite{luca},
the most general ELKO matter fields including the dynamical torsion was constructed,
whose cosmology leads to anisotropic universes.
To the knowledge of the authors, the relation between the two approaches has not been explored, and  it seems that at present only  the former approach could provide the novel  cosmological feature of  addressing the coincidence problem for the dark spinor.

One interesting feature of the potential (\ref{ppp}) is that in contrast with the pure exponential potential in the quintessence model
where the parameter $\lambda$ is restricted to be within a certain range to produce the attractor solution,
the solution with the torsion is an attractor for each value of $\lambda>0$.
Moreover, the term $e^{-\lambda\phi/2}\phi^2$ is reminiscent of the Albrecht-Skordis potential \cite{albr} which appears in the low energy limit of $M$ theory.
The presence of the torsion is a key ingredient and possible implications of this aspect deserve further investigations.


We conclude with the following
 remarks. As shown in Eq. (\ref{massy}) of Sec. IV B, there is a relation between the cosmological constant
and the mass of the cosmological particle $\phi$ in our dark spinor model with torsion,
which might represent the main component of our present universe and is about
$ 10^{-63} g $, as in the massive vector model \cite{mass}.
When the analysis is extended to the case including a conformal coupling of dark spinor fields to gravity, it will provide with a
richer variety of cosmological consequences
than the case without torsion \cite{qc}. Last but not least, possible implications of the experimental limits on torsion \cite{russel} in the dark spinor case need to be explored.


\acknowledgements

T. H. L. was supported by the Basic Science Research
Program through the National Research Foundation of
Korea (NRF) funded by Ministry of Education, Science
and Technology (2012-0003008). P. O. was supported by
the Basic Science Research Program through the National
Research Foundation of Korea (NRF) funded by Ministry of
Education, Science and Technology (2010-0021996) and
by a NRF grant funded by the Korean government (MEST)
through the Center for Quantum Spacetime (CQUeST) of
Sogang University with Grant No. 2005-0049409.

\end{document}